# Asteroids Fail to Retain Cometary Impact Signatures

Sarah Joiret[1], Guillaume Avice[2], Ludovic Ferrière[3,4], Zoë M. Leinhardt[5], Simon Lock[6], Alexandre Mechineau[1], and Sean N. Raymond[1]
[1] Laboratoire d'Astrophysique de Bordeaux, Univ. Bordeaux, CNRS, B18N, allée Geoffroy Saint-Hilaire, 33615 Pessac, France
[2] Université Paris Cité, Institut de physique du globe de Paris, CNRS, 75005 Paris, France
[3] Natural History Museum Vienna, Burgring 7, A-1010 Vienna, Austria
[4] Natural History Museum Abu Dhabi, Saadiyat Island, Abu Dhabi, UAE
[5] H.H. Wills Physics Laboratory, School of Physics, University of Bristol, Bristol, UK
[6] School of Earth Sciences, University of Bristol, Bristol, UK

## Abstract

A bombardment of comets is thought to have occurred in the inner solar system as a result of a dynamical instability among the giant planets after gas disk dispersal. Vesta, the second largest asteroid in the main asteroid belt, likely differentiated before gas disk dispersal, implying its crust witnessed the cometary bombardment. The composition of howardite, eucrite, and diogenite (HED) meteorites, which represent fragments of Vesta's crust, could therefore have been altered by cometary impacts. Here we combine noble gas mass spectrometry measurements, *N*-body simulations, collision rate calculations, and impact simulations to estimate the cometary contribution to Vesta. While our dynamical simulations indicate that Vesta likely underwent a significant number of collisions with large comets, we find no xenon cometary signature in HED meteorites. This apparent contradiction arises due to the fact that cometary impacts were at high speeds and Vesta's weak gravitational attraction made it incapable of retaining cometary material. Smaller asteroids are even less likely to retain such material. Therefore, if a cometary xenon signature is ever detected in an asteroid belt object, it must have been acquired during formation, within the same source region as comet 67P/Churyumov-Gerasimenko, and have been implanted later into the asteroid belt.

*Unified Astronomy Thesaurus concepts:* Small Solar System bodies (1469); Asteroids (72); Comet dynamics (2213); Isotopic abundances (867); Solar system evolution (2293); Solar system (1528)

## 1. Introduction

The current architecture of the solar system was largely shaped by the formation and early evolution of the giant planets. The formation of the giant planets led to a first bombardment of asteroids to the inner solar system (S. N. Raymond & A. Izidoro 2017a). Comets were not significantly delivered to the inner solar system at this stage. A later, more disruptive event—the giant planet instability—occurred after gas disk dispersal and scattered comets to the inner and outer solar system (R. Gomes et al. 2005; K. Tsiganis et al. 2005; D. Nesvorný 2018), while also shifting the locations of mean motion and secular resonances across the asteroid belt (D. A. Minton & R. Malhotra 2011). During this second, more intense phase of bombardment, both asteroids and comets impacted the inner planetary bodies, likely contributing to their volatile inventories (B. Marty et al. 2016, 2017). The timing of the giant planet instability remains uncertain, but it is thought to have occurred within the first 100 million years after gas disk dispersal, possibly even immediately after (A. Morbidelli et al. 2018; D. Nesvorny et al. 2018; S. J. Mojzsis et al. 2019; B. Quarles & N. Kaib 2019; R. d. S. Ribeiro et al. 2020; A. C. Hunt et al. 2022; B. Liu et al. 2022; G. H. Edwards et al. 2024). To understand solar system formation it is necessary to determine the effect that this instability—following which collision probabilities with comets were highest in the inner solar system—had on the terrestrial planets and asteroids.

Xenon has emerged as an effective tool for discriminating chondritic and cometary contributions to planetary bodies, owing to the distinctive isotopic signatures of chondrites and the recent first measurement from comet 67P/Churyumov-Gerasimenko (D. V. Bekaert et al. 2017; B. Marty et al. 2017; J. F. Pernet-Fisher et al. 2020; X. Zhang et al. 2024). This measurement is thought to be at least somewhat representative of a population of comets because its isotopic characteristics—particularly the deficits in heavy isotopes—align with long-standing expectations for cometary material inferred from the unexplained features of Earth's primordial atmospheric xenon (B. Marty et al. 2016). Indeed, the xenon signature in Earth's primordial atmosphere, known as U-Xe (R. O. Pepin 1991; R. Pepin 1994), is best interpreted as a mixture comprising approximately 22% cometary (67P/C-G) xenon and 78% chondritic xenon (B. Marty et al. 2017). Consequently, the identification of a U-Xe-like signature in a planetary object can be reasonably taken as evidence of cometary contribution.

Noble gas isotopic analysis can also be applied to understand impact histories on other planetary bodies, such as Vesta. Howardite, eucrite, and diogenite (HED) meteorites are thought to originate from Vesta's crust (T. B. McCord et al. 1970; G. J. Consolmagno & M. J. Drake 1977; M. C. De Sanctis et al. 2012). These three types of meteorites constitute the largest sampling available for any differentiated asteroid (H. Y. McSween et al. 2011). Their formation ages limit the completion of Vesta's core–mantle differentiation and crust crystallization to 1 and 3 Myr after the condensation of calcium–aluminum rich inclusions (CAIs), respectively

(M. Bizzarro et al. 2005; M. Schiller et al. 2011; M. Touboul et al. 2015; G. Hublet et al. 2017). Because the differentiation likely occurred prior to gas disk dispersal, estimated at <4–5 Myr after CAIs in the case of the solar system (H. Wang et al. 2017; B. P. Weiss & W. F. Bottke 2021; A. C. Hunt et al. 2022), Vesta's crust should have witnessed the bombardment caused by the giant planet instability. In previous work, the highly siderophile element budget in HED meteorites was used to infer the intensity and evolution of the chondritic impact rate on Vesta since core–mantle differentiation (D. Turrini et al. 2018; M.-H. Zhu et al. 2021). Similarly, their xenon isotopic signature could, in principle, record the cometary impact rate on their parent body (B. Marty et al. 2017), if this cometary signature can be retained. In the 1990s, a detection of U-Xe was claimed in the diogenite Tatahouine (T. Michel & O. Eugster 1994). However, later measurements on the same meteorite refuted these results (H. Busemann & O. Eugster 2002). Finding a U-Xe signature, or any xenon signature comprising a cometary component, in HED meteorites would provide evidence of the cometary bombardment on Vesta. We note that Ceres—the largest object in the asteroid belt—has not yet been linked to any known meteorite group, limiting our ability to apply similar geochemical approaches to its history.

Vesta is thought to have formed in the terrestrial planet region before being implanted into the asteroid belt early in its history (W. F. Bottke et al. 2006; S. N. Raymond & A. Izidoro 2017b; A. Mastrobuono-Battisti & H. B. Perets 2017; M.-H. Zhu et al. 2021; R. Deienno et al. 2024, 2025). Vesta is part of an asteroid family—Vestoids—that share similar orbits and are thought to have formed from the fragmentation of a common parent body after two major impact events (F. Marzari et al. 1996; P. Schenk et al. 2012). Because the Vesta family is relatively tightly clustered in the asteroid belt, the events that disrupted the Vestoids parent body likely occurred after the giant planet instability and main bombardment episodes (D. Nesvorný et al. 2008; P. I. O. Brasil et al. 2016). In addition, Vesta experienced major impacts between 4.1 and 3.5 Ga, as recorded by $^{40}$Ar-$^{39}$Ar shock ages of HEDs (D. Bogard & D. Garrison 2009), that could have produced the Vestoids (H. Y. McSween et al. 2011; F. Spoto et al. 2015). These events are therefore unrelated to the dynamical instability and cometary bombardment. However, it is reasonable to expect that if Vesta exhibits a cometary signature as a result of the instability, Vestoids, and in turn HEDs, should also carry this signature.

Here we investigated the effect of cometary impacts on Vesta by measuring the xenon isotopic composition of two HED meteorites, the diogenites Tatahouine and Shalka, in order to independently confirm the findings of H. Busemann & O. Eugster (2002). Diogenites are well suited to study the origin of xenon in HED meteorites, as they contain lower concentrations of cosmogenic and fissiogenic xenon than eucrites (T. Michel & O. Eugster 1994). We combined these measurements with *N*-body simulations, collision rate calculations, and impact simulations of comets on a Vesta-like body, addressing the efficiency of cometary material retention.

## 2. Methods

### 2.1. Noble Gas Mass Spectrometry

Noble gases were extracted from the diogenite samples by stepwise heating in an induction furnace (see results in Tables 1 and 2). This method enables the separation of noble gas components from each other. The extracted gas is purified from all the chemically active species through a purification line consisting of several getters in series. Separation of noble gases is achieved through cryogenic trapping on charcoal. Hence, each noble gas is sequentially released by heating the charcoal with a resistor and admitted to the mass spectrometer separately. Data were corrected for blank contribution and mass discrimination of the instrument through standard measurements. Additional details can be found in G. Avice et al. (2023).

In our figures we use the delta notation. It refers to the relative difference, in parts per thousand, of the isotopic ratios of a sample and of a reference standard, e.g., air. For example, with terrestrial atmospheric Xe as the reference, we have

$$\delta^i\mathrm{Xe}_{\mathrm{air}} = \left( \frac{\frac{^i\mathrm{Xe}}{^{132}\mathrm{Xe}}}{\left(\frac{^i\mathrm{Xe}}{^{132}\mathrm{Xe}}\right)_{\mathrm{air}}} - 1 \right) \times 1000.$$

On a plot of $\delta^i\mathrm{Xe}_{132}$ relative to $i$, the line $y = 0$ corresponds to atmospheric Xe.

### 2.2. *N*-body Simulations

Simulation outputs from S. Joiret et al. (2024) were used to calculate the collision rate of comets with a Vesta-like asteroid. These *N*-body simulations of the solar system were carried out using GENGA, a hybrid symplectic integrator. The different sets of simulations were started at the time of gas disk dispersal ($t_0$) and lasted for 100 Myr. Only gravitational forces were considered. Planetary embryos and larger bodies were in full gravity mode, whereas smaller bodies such as planetesimals and comets could only interact with larger bodies.

At the moment of gas disk dispersal ($t_0$), i.e., at the start of our simulations, the giant planets were already formed. The inner solar system, on the other hand, comprised a bimodal population of planetesimals and Moon- to Mars-sized embryos, uniformly distributed into a ring centered between Venus's and Earth's orbits. The total mass of the ring was set to 2.5 $M_{\mathrm{Earth}}$, comprising 35 embryos of $\simeq 1/2$ of Mars's mass and 2000 planetesimals of $1.25 \times 10^{-4}\ M_{\mathrm{Earth}} \simeq 10^{-2}\ M_{\mathrm{Moon}}$ each (that is, about the mass of Ceres). These embryos and planetesimals were assumed to be entirely noncarbonaceous. We considered 1600 carbonaceous planetesimals with the exact same mass as the noncarbonaceous ones, amounting to a total of 0.2 $M_{\oplus}$ (S. N. Raymond & D. Nesvorný 2022).

We also considered an outer ring of cometary planetesimals uniformly distributed between 21 and 30 au, with a total mass of 25 $M_{\mathrm{Earth}}$. This configuration is suggested by models of the giant planet instability (D. Nesvorný 2018) accounting for losses by collisional evolution within the primordial reservoir (W. F. Bottke et al. 2023). The outer ring consisted of $10^4$ comets with $2.5 \times 10^{-3}\ M_{\mathrm{Earth}} \simeq 1.5 \times 10^{25}\ \mathrm{g} \simeq M_{\mathrm{Pluto}}$. It was shown that the primordial outer disk of comets should have had some hundreds to thousands of Pluto-mass bodies, and they could have represented 10%–40% of the total disk mass (D. Nesvorný & D. Vokrouhlický 2016; N. A. Kaib et al. 2024). Here, they represented 100% of the total disk mass. Indeed, taking into account the effect of smaller comets would have implied a need to increase the total number of particles in our simulation for a fixed outer disk mass, and in turn, to increase the computational costs associated with our simulations. We included the early instability model (M. Clement et al. 2018) by forcing the giant planets into a dynamical instability ∼2 Myr after gas disk dispersal. This was done using simulation outputs from M. S. Clement et al. (2021b) and the `set_elements` option in GENGA, which enables bodies within a simulation to follow a precomputed orbital evolution. More details about our *N*-body simulations can be found in S. Joiret et al. (2024).

**Table 1**
Xe Concentrations and Isotopic Ratios in Diogenite Tatahouine for Various Heating Steps

| Temperature (°C) | $^{130}$Xe (ccSTP g$^{-1}$) | $\frac{^{124}Xe}{^{130}Xe}$ | $\frac{^{126}Xe}{^{130}Xe}$ | $\frac{^{128}Xe}{^{130}Xe}$ | $\frac{^{129}Xe}{^{130}Xe}$ | $\frac{^{131}Xe}{^{130}Xe}$ | $\frac{^{132}Xe}{^{130}Xe}$ | $\frac{^{134}Xe}{^{130}Xe}$ | $\frac{^{136}Xe}{^{130}Xe}$ |
|---|---|---|---|---|---|---|---|---|---|
| 850 | 1.47E-13 | 0.0613 | 0.0259 | 0.4712 | 6.482 | 5.586 | 6.655 | 2.521 | 2.158 |
| | ±5E-15 | ±0.0030 | ±0.0013 | ±0.0058 | ±0.061 | ±0.063 | ±0.052 | ±0.023 | ±0.020 |
| 1185 | 2.79E-13 | 0.0744 | 0.0272 | 0.4726 | 6.515 | 6.174 | 6.707 | 2.536 | 2.187 |
| | ±1.0E-14 | ±0.0034 | ±0.0011 | ±0.0050 | ±0.055 | ±0.058 | ±0.044 | ±0.020 | ±0.017 |
| 1485 | 8.44E-13 | 0.0261 | 0.0221 | 0.4679 | 6.488 | 5.227 | 6.507 | 2.515 | 2.145 |
| | ±3.0E-14 | ±0.0012 | ±0.0008 | ±0.0044 | ±0.052 | ±0.044 | ±0.039 | ±0.018 | ±0.016 |
| 1622 | 1.84E-12 | 0.0213 | 0.0217 | 0.4671 | 6.479 | 5.148 | 6.511 | 2.510 | 2.146 |
| | ±6E-14 | ±0.0009 | ±0.0008 | ±0.0042 | ±0.051 | ±0.043 | ±0.038 | ±0.018 | ±0.016 |
| 1700 | 9.06E-13 | 0.0225 | 0.0222 | 0.4694 | 6.490 | 5.181 | 6.529 | 2.519 | 2.147 |
| | ±3.3E-14 | ±0.0010 | ±0.0009 | ±0.0045 | ±0.052 | ±0.044 | ±0.040 | ±0.019 | ±0.016 |
| 1870 | 8.08E-13 | 0.0219 | 0.0212 | 0.4669 | 6.457 | 5.131 | 6.510 | 2.507 | 2.135 |
| | ±2.9E-14 | ±0.0010 | ±0.0008 | ±0.0045 | ±0.053 | ±0.044 | ±0.041 | ±0.019 | ±0.016 |
| 1930 | 3.57E-13 | 0.0206 | 0.0213 | 0.4677 | 6.529 | 5.161 | 6.489 | 2.516 | 2.142 |
| | ±1.3E-14 | ±0.0011 | ±0.0010 | ±0.0054 | ±0.058 | ±0.048 | ±0.047 | ±0.021 | ±0.018 |

**Table 2**
Xe Concentrations and Isotopic Ratios in Diogenite Shalka for Various Heating Steps

| Temperature (°C) | $^{130}$Xe (ccSTP g$^{-1}$) | $\frac{^{124}Xe}{^{130}Xe}$ | $\frac{^{126}Xe}{^{130}Xe}$ | $\frac{^{128}Xe}{^{130}Xe}$ | $\frac{^{129}Xe}{^{130}Xe}$ | $\frac{^{131}Xe}{^{130}Xe}$ | $\frac{^{132}Xe}{^{130}Xe}$ | $\frac{^{134}Xe}{^{130}Xe}$ | $\frac{^{136}Xe}{^{130}Xe}$ |
|---|---|---|---|---|---|---|---|---|---|
| 850 | 8.43E-13 | 0.1904 | 0.0346 | 0.4678 | 6.421 | 8.214 | 6.852 | 2.600 | 2.383 |
| | ±3.0E-14 | ±0.0083 | ±0.0013 | ±0.0044 | ±0.052 | ±0.071 | ±0.042 | ±0.019 | ±0.018 |
| 1185 | 1.77E-12 | 0.1189 | 0.0283 | 0.4671 | 6.458 | 6.678 | 6.719 | 2.577 | 2.309 |
| | ±6E-14 | ±0.0052 | ±0.001 | ±0.0042 | ±0.051 | ±0.057 | ±0.040 | ±0.018 | ±0.017 |
| 1485 | 1.99E-12 | 0.0242 | 0.0219 | 0.4668 | 6.493 | 5.230 | 6.534 | 2.517 | 2.142 |
| | ±7E-14 | ±0.0011 | ±0.0008 | ±0.0042 | ±0.051 | ±0.043 | ±0.038 | ±0.018 | ±0.016 |
| 1622 | 7.36E-13 | 0.0217 | 0.0218 | 0.4685 | 6.379 | 5.089 | 6.479 | 2.515 | 2.157 |
| | ±2.7E-14 | ±0.0011 | ±0.0007 | ±0.0073 | ±0.108 | ±0.084 | ±0.102 | ±0.038 | ±0.040 |
| 1800 | 1.49E-13 | 0.0276 | 0.0273 | 0.4755 | 6.584 | 5.258 | 6.614 | 2.601 | 2.332 |
| | ±5E-15 | ±0.0018 | ±0.0016 | ±0.0102 | ±0.126 | ±0.099 | ±0.121 | ±0.047 | ±0.049 |
| 1930 | 1.36E-13 | 0.0201 | 0.0186 | 0.4715 | 6.602 | 5.176 | 6.576 | 2.540 | 2.208 |
| | ±5E-15 | ±0.0015 | ±0.0013 | ±0.0110 | ±0.130 | ±0.101 | ±0.123 | ±0.047 | ±0.047 |

Out of 20 simulations, 10 were considered successful because they included realistic Venus, Earth, and Mars analogs by the end of the integration (see Section 2.1.2. in S. Joiret et al. 2024 for the criteria used to define successful Venus, Earth, and Mars analogs). Among these 10 simulations, we found 25 planetesimals that originated in the terrestrial planet region and ended up with a semimajor axis between 2 and 3 au. Of these, eight achieved a stable orbit by the end of the simulation (i.e., $e \lesssim 0.3$ and $i \lesssim 20°$). These stable planetesimals were classified as Vesta-like, with an average of one Vesta-like planetesimal in each simulation.

### 2.3. Collision Rate Calculations

For each time step of each simulation comprising a Vesta-like asteroid, we calculated the relative velocity and collision probability between each of the $10^4$ comets and the Vesta-like asteroid, and each of the 1600 carbonaceous asteroids and the Vesta-like asteroid. They were obtained using an Öpik/Wetherill method (G. W. Wetherill 1967) including considerations from P. Farinella & D. R. Davis (1992). Details of this algorithm can be found in S. Joiret et al. (2023). Additionally, we incorporated considerations from W. F. Bottke et al. (1994) to weight the collision velocities by the collision probability per unit of time at each orientation. This allowed us to refine the assessment of collision velocities for each pair of bodies.

The relative velocity $U$ between a comet and a Vesta-like asteroid is

$$U = \frac{|U_+| + |U_-|}{2} \quad (1)$$

expressed in km yr$^{-1}$, where

$$U_+^2 = \frac{GM}{r}\left[4 - \frac{r}{a_V} - \frac{r}{a_c} - 2\left[\frac{a_V a_c}{r^2}(1 - e_V^2) \times (1 - e_c^2)\right]^{\frac{1}{2}}(\cos i' + |\cot \alpha_V||\cot \alpha_c|)\right], \quad (2)$$

$$U_-^2 = \frac{GM}{r}\left[4 - \frac{r}{a_V} - \frac{r}{a_c} - 2\left[\frac{a_V a_c}{r^2} \times (1 - e_V^2)(1 - e_c^2)\right]^{\frac{1}{2}}(\cos i' - |\cot \alpha_V||\cot \alpha_c|)\right]. \quad (3)$$

$GM$ is the heliocentric gravitational constant and is equal to $1.3202 \times 10^{26}\,\mathrm{km^3\,kg^{-1}\,yr^{-2}}$; $a_V$ and $a_c$ are the semimajor axes of the Vesta-like asteroid and the comet, respectively; $e_V$ and $e_c$ are the eccentricity of the Vesta-like asteroid and the comet, respectively; $r$ is the distance of the Vesta-like asteroid from the Sun; $i'$ is the mutual inclination of the two bodies; and

$$\cot\alpha_V = \pm\sqrt{\frac{a_V^2 e_V^2 - r^2\left(\frac{a_V}{r} - 1\right)^2}{a_V^2(1 - e_V^2)}}, \tag{4}$$

$$\cot\alpha_c = \pm\sqrt{\frac{a_c^2 e_c^2 - r^2\left(\frac{a_c}{r} - 1\right)^2}{a_c^2(1 - e_c^2)}}. \tag{5}$$

$U$ should be integrated over all possible true anomalies of the Vesta-like asteroid $\theta_V$, and over all possible relative longitudes of the ascending node of the two bodies $\Delta\Omega$. We then divided it by $3.154 \times 10^7$ to have it expressed in $\mathrm{km\,s^{-1}}$.

The intrinsic collision probability per unit of time between a Vesta-like asteroid and a comet is

$$P_i = \frac{(|U_+| + |U_-|)\, r\, c}{|\cot\alpha_c|}, \tag{6}$$

where

$$c = \frac{1}{8\pi^2 \sin i' a_V^2 a_c^2 \sqrt{(1 - e_V^2)(1 - e_c^2)}}. \tag{7}$$

$P_i$ must be integrated over all $\theta_V$, and $\Delta\Omega$ to obtain the total intrinsic collision probability per unit of time $P_{\mathrm{tot}}$. The latter defines the collision probability per unit of time between two bodies when the sum of their respective radii equals 1 km.

To account for W. F. Bottke et al.'s (1994) method of weighting collision velocities by collision probability, we used a weighted velocity $U_{\mathrm{weighted}}$ calculated as follows:

$$U_{\mathrm{weighted}} = \frac{\int_{\Delta\Omega}\int_{\theta_V} P_i(\theta_V, \Delta\Omega)\, U(\theta_V, \Delta\Omega)\, d\theta_V \quad d(\Delta\Omega)}{\int_{\Delta\Omega}\int_{\theta_V} P_i(\theta_V, \Delta\Omega)\, d\theta_V \quad d(\Delta\Omega)}. \tag{8}$$

Finally, the number of collisions between a Vesta-like asteroid and a comet occurring during a certain time interval $\Delta t$ can be expressed as

$$n_{\mathrm{coll}} = P_{\mathrm{tot}}(R + r)^2 f_{\mathrm{grav}}\,\Delta t, \tag{9}$$

where $R$ and $r$ are the radii in kilometers of the Vesta-like asteroid and the comet, respectively, and $f_{\mathrm{grav}}$ (dimensionless) is the gravitational focus of the Vesta-like asteroid given by

$$f_{\mathrm{grav}} = 1 + \frac{v_{\mathrm{esc},V}^2}{U^2} \tag{10}$$

with $v_{\mathrm{esc},V} = \sqrt{\frac{2GM_V}{R}}$ the escape velocity of the Vesta-like asteroid.

We added up the contribution of all comets (i.e., added up the $n_{\mathrm{coll}}$ of all comets at each time step) and normalized this value to the total number of comets in our simulations (i.e., divided it by $10^4$). We thereby obtained the number of collisions per comet in the primordial outer disk of comets, during a certain time interval.

For each Vesta-like asteroid, we considered all comets and carbonaceous asteroids with a nonzero collision probability, and calculated the mean and minimum collision velocity (weighted by the collision probability) between all pairs of bodies at each time step.

### 2.4. Impact Simulations

We used the shock physics code iSALE2D (version 4.1) to estimate the efficiency of material accretion depending on the velocity of cometary impactors. The iSALE shock physics code (K. Wünnemann et al. 2006) is an extension of the SALE hydrocode (A. A. Amsden et al. 1980). To simulate hypervelocity impact processes in solid materials, SALE was modified to include an elastoplastic constitutive model, fragmentation models, various equations of state (EOS), and multiple materials (H. J. Melosh et al. 1992; B. A. Ivanov et al. 1997). More recent improvements include a modified strength model (G. S. Collins et al. 2004; K. Wünnemann et al. 2006), a porosity compaction model (G. S. Collins et al. 2011), and a dilatancy model (G. S. Collins 2014).

Asteroids were assumed to be dunitic, modeled by the ANEOS (using the default parameters in iSALE2D). Comets were assumed to be homogeneous bodies made of pure water ice, modeled by the Tillotson EOS (also using the default parameters in iSALE2D). In reality, they are a mixture of approximately half ices and half silicates and refractory organic materials (J. M. Greenberg 1998). D. Turrini et al. (2018) explored retention efficiencies for comets made of water ice mixed with a rocky component, on Vesta. However, water ice remains the most suitable proxy for cometary material in impact simulations using the EOS currently available in iSALE2D, given that comets are thought to have a lower density than water (J. E. Richardson et al. 2007; O. Groussin et al. 2019). So the modeled impacts in this study, and in that of D. Turrini et al. (2018), likely overestimate the fraction of projectile material that would actually accrete on a Vesta-like asteroid. Furthermore, laboratory impact experiments show that projectile retention is higher for higher impact angle (measured from horizontal; R. T. Daly & P. H. Schultz 2016). Consequently, vertical water-ice impactors modeled in this study represent an upper limit of the projectile material that may be accreted on a Vesta-like asteroid.

Cometary impactors were resolved by 50 cells per projectile radius, where each cell had a side length of 100 m, resulting in a radius of 5 km. Vesta's crust was modeled by the ANEOS (using the default parameters in iSALE2D) as a 50 km thick basaltic layer. The gravitational acceleration and radius of the target were set to $g_{\mathrm{Vesta}} = 0.22\,\mathrm{m\,s^{-2}}$ and 260 km, respectively. The computational domain was a 2D box with cylindrical symmetry, comprising 400 high-resolution cells along the horizontal axis and 300 high-resolution cells along the vertical axis, corresponding to physical dimensions of $80 \times 30$ km. We chose free-slip for the left and right boundary conditions, no-slip for the bottom, and outflow for the top. All simulations began with the first contact between the impactor and the top of the atmosphere and stopped after the collapse of the transient crater.

We find that for head-on impactors with a diameter of 10 km, 100% of the material reaches escape velocity for cometary impacts at $\geqslant 5\,\mathrm{km\,s^{-1}}$ and for asteroidal impacts at $\geqslant 14\,\mathrm{km\,s^{-1}}$. We note that, following the giant planet instability, when collision probabilities with comets were highest in the inner solar system, collision velocities were never as low as 5 km s$^{-1}$ and only on extremely rare occasions could have been as low as 10 km s$^{-1}$. In contrast, asteroidal impactors could frequently have velocities below 14 km s$^{-1}$, allowing for partial retention of material.

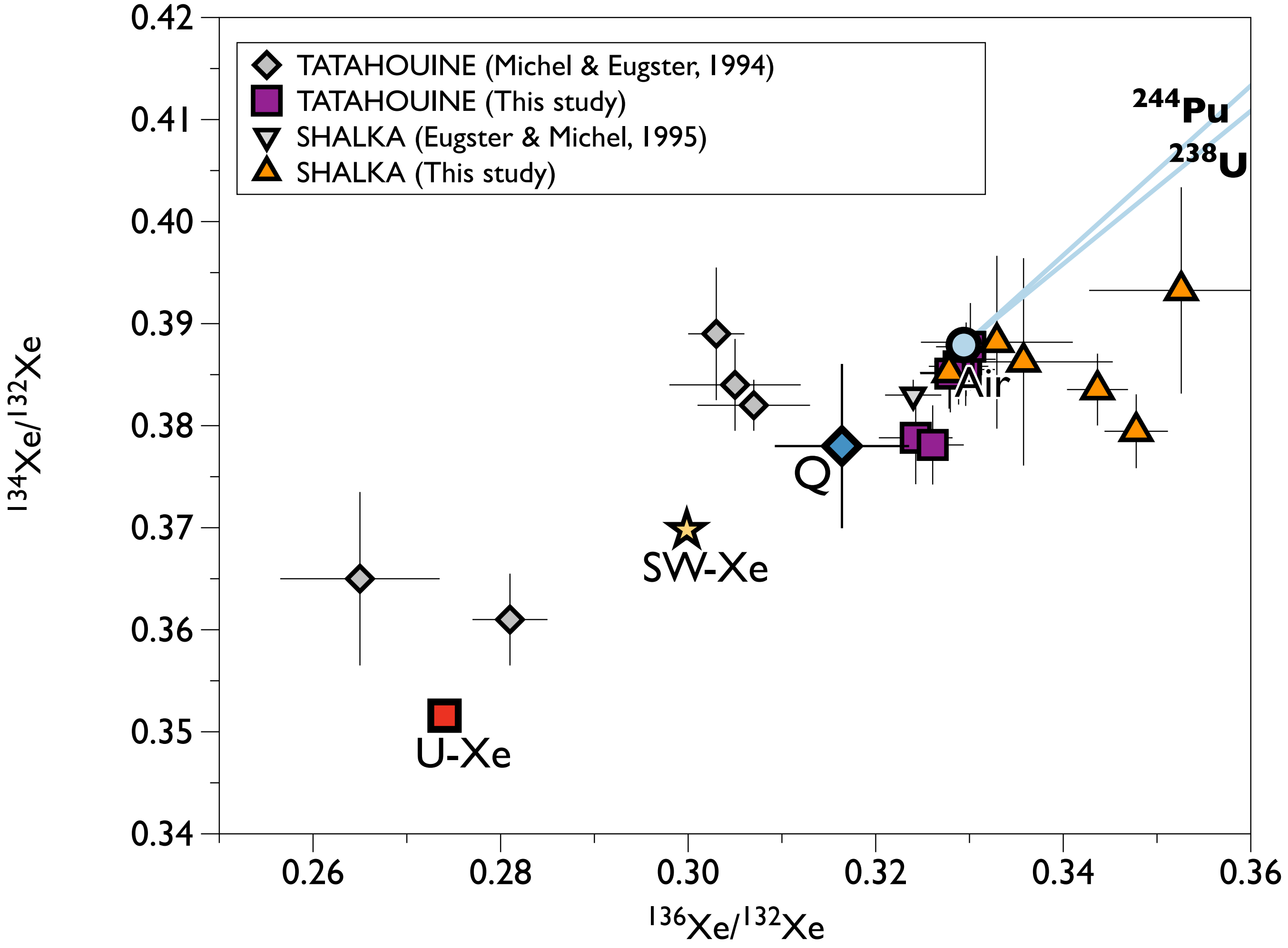

**Figure 1.** The $^{134}Xe/^{132}Xe$ vs. $^{136}Xe/^{132}Xe$ diagram showing the composition of Tatahouine (#13349) and Shalka (#6766_C) measured at different heating steps, including those of T. Michel & O. Eugster (1994) and O. Eugster & T. Michel (1995) for comparison. The $^{238}U$ and $^{244}Pu$ fission Xe data are from D. Porcelli et al. (2002). Q-Xe (Q phase is the main carrier of heavy noble gases in chondrites) is from H. Busemann et al. (2000), SW-Xe (SW stands for solar wind) is from A. Meshik et al. (2020), U-Xe is from R. O. Pepin & D. Porcelli (2002), and Air (corresponding to Earth's atmosphere) is from M. Ozima & F. A. Podosek (2001). Error bars are at 1$\sigma$.

## 3. Experimental Results

We did not find any U-Xe or other cometary signature in the investigated samples of Tatahouine (#13349) and Shalka (#6766_C), both from the Natural History Museum Vienna collection. We compared our results with those of T. Michel & O. Eugster (1994; hereafter ME94), where the Tatahouine measurements indicated a U-Xe composition released during the 1000°C and 1200°C heating steps. Unlike ME94, we found that the major fraction of Xe trapped in Tatahouine is of terrestrial atmospheric composition, without a U-Xe component, at any heating step (Figure 1). We also found an atmospheric composition for Shalka, similarly to O. Eugster & T. Michel (1995).

Figure 2 shows the isotopic pattern of Xe in Tatahouine and Shalka, relative to terrestrial atmospheric Xe. We compared our results to other studies (T. Michel & O. Eugster 1994; O. Eugster & T. Michel 1995), and included those of H. Busemann & O. Eugster (2002; hereafter BE02), who also found an atmospheric composition for Tatahouine with no U-Xe component. The black squares indicate the isotopic composition of U-Xe for reference. While the results of ME94 appear to follow the characteristic U-Xe trend, our measurements, consistent with those of BE02, deviate from this pattern and instead align closely with a purely atmospheric xenon signature, thereby providing no evidence for a cometary contribution.

The reason for the discrepancy between ME94's results and those of subsequent studies (including BE02 and ours) lacks clear explanation. BE02 argued that ME94 was incorrect on the grounds of physical plausibility, as it appears unlikely for achondrites (comprising HEDs), formed by crystallization of magma, to contain primordial noble gases, such as U-Xe. Indeed, differentiation of Vesta at temperatures of ~1600°C should have led to a significant loss of volatiles, as indicated by the generally small concentrations of trapped Xe in achondrites relative to chondrites. These low Xe abundances could easily be overprinted by atmospheric Xe contamination. However, it is now commonly accepted that a U-Xe composition is linked to a cometary contribution

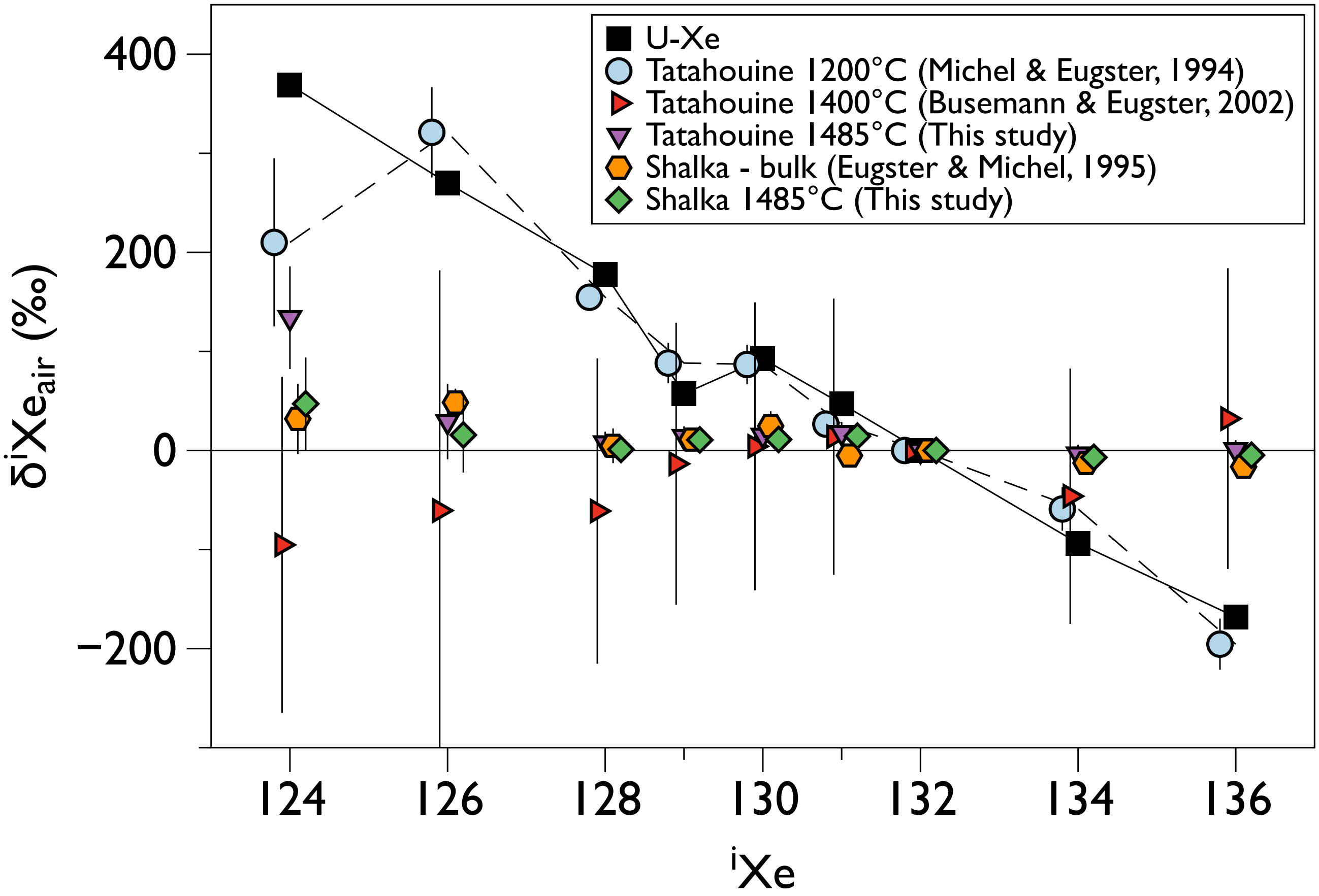

**Figure 2.** Measured isotopic pattern of Xe in Tatahouine and Shalka compared to T. Michel & O. Eugster (1994), O. Eugster & T. Michel (1995), and H. Busemann & O. Eugster (2002) at specific heating steps. U-Xe data are from R. O. Pepin & D. Porcelli (2002). The $\delta$ notation refers to the relative difference, in parts per thousand, of the isotopic ratios of a sample and of a reference standard. Using atmospheric Xe as the reference, $\delta^i\mathrm{Xe}_{\mathrm{air}} = \left( \frac{\frac{^i\mathrm{Xe}}{^{132}\mathrm{Xe}}}{\left(\frac{^i\mathrm{Xe}}{^{132}\mathrm{Xe}}\right)_{\mathrm{air}}} - 1 \right) \times 1000$. The horizontal line at $y = 0$ represents terrestrial atmospheric Xe. Error bars indicate $\pm 1\sigma$.

(B. Marty et al. 2017), probably as a result of the cometary bombardment, which occurred after Vesta's differentiation. Therefore, it is unlikely that U-Xe was hidden due to the differentiation processes on Vesta, as suggested by BE02. Yet the atmospheric contamination hypothesis remains plausible. Achondrites in general have proven to trap significant amounts of air during weathering processes (S. Niedermann & O. Eugster 1992). And BE02 found that the samples of Tatahouine crushed to powder contain 1 order of magnitude less atmospheric Xe than the less fine-grained samples, suggesting that the terrestrial contamination was introduced prior to the sample preparation.

We argue that the results of ME94 are implausible, not because of the differentiation processes hiding the cometary signature, as proposed by BE02, but because it is unlikely that such a signature could be efficiently retained on asteroids the size of Vesta or smaller, as demonstrated in the following section.

## 4. Simulation Results

We investigated the dynamical implantation of Vesta into the asteroid belt using results from previous dynamical $N$-body simulations of the formation of the solar system (S. Joiret et al. 2024). These simulations included terrestrial planet-forming embryos and planetesimals, carbonaceous asteroids, comets, and giant planets following prescribed orbital pathways to evolve into a dynamical instability shortly after gas disk dispersal. We found approximately one Vesta-like planetesimal in each simulation. Of eight Vesta analogs, three were implanted $\sim$20 Myr after $t_0$ (the time of gas disk dispersal), two were implanted $\sim$40 Myr after $t_0$, and three were implanted >60 Myr after $t_0$ (Figure 3, lower panel).

In each simulation with a Vesta analog, we calculated the collision probabilities between these Vesta analogs and comets. Within the first 100 Myr after gas disk dispersal, the cumulative collision rate of comets with a Vesta-like asteroid ranges between $7 \times 10^{-8}$ and $2 \times 10^{-7}$ per comet, depending on the simulation (Figure 3, upper panel). Considering that there were $\simeq 5 \times 10^7$ $D > 100$ km objects and $\simeq 10^{10}$ $D > 10$ km objects in the primordial outer disk of comets (D. Nesvorný et al. 2023), a collision between Vesta and a >100 km comet is plausible, and collisions between Vesta and >10 km comets were probably numerous. Geological evidence for these cometary collisions may be challenging to identify on Vesta, as they were later overprinted by much larger impacts that formed the two major impact basins (F. Marzari et al. 1996; P. Schenk et al. 2012).

Figure 3 shows the dynamical evolution of Vesta-like objects in the context of the giant planet instability (bottom panel), and the collision rates of comets on these Vesta-like objects through time (upper panel). It reveals three notable features. First, the implantation of planetesimals from the terrestrial planet-forming region to the asteroid belt may occur tens of millions of years after the timing of the giant planet instability. This is consistent with A. Izidoro et al.'s (2024) proposal that the timing of such asteroid implantation does not constrain the timing of the giant planet instability. Second, most cometary impacts occurred prior to the implantation of Vesta-like bodies into the asteroid belt and, in any case, took place very early in solar system history. This timing suggests that cometary impacts were unrelated to the subsequent disruption of the Vestoid parent body, which is inferred to have occurred later based on the tight orbital clustering of the Vestoids (D. Nesvorný et al. 2008). Third, the absence of cometary Xe within HED meteorites cannot be attributable to a lack of cometary impacts.

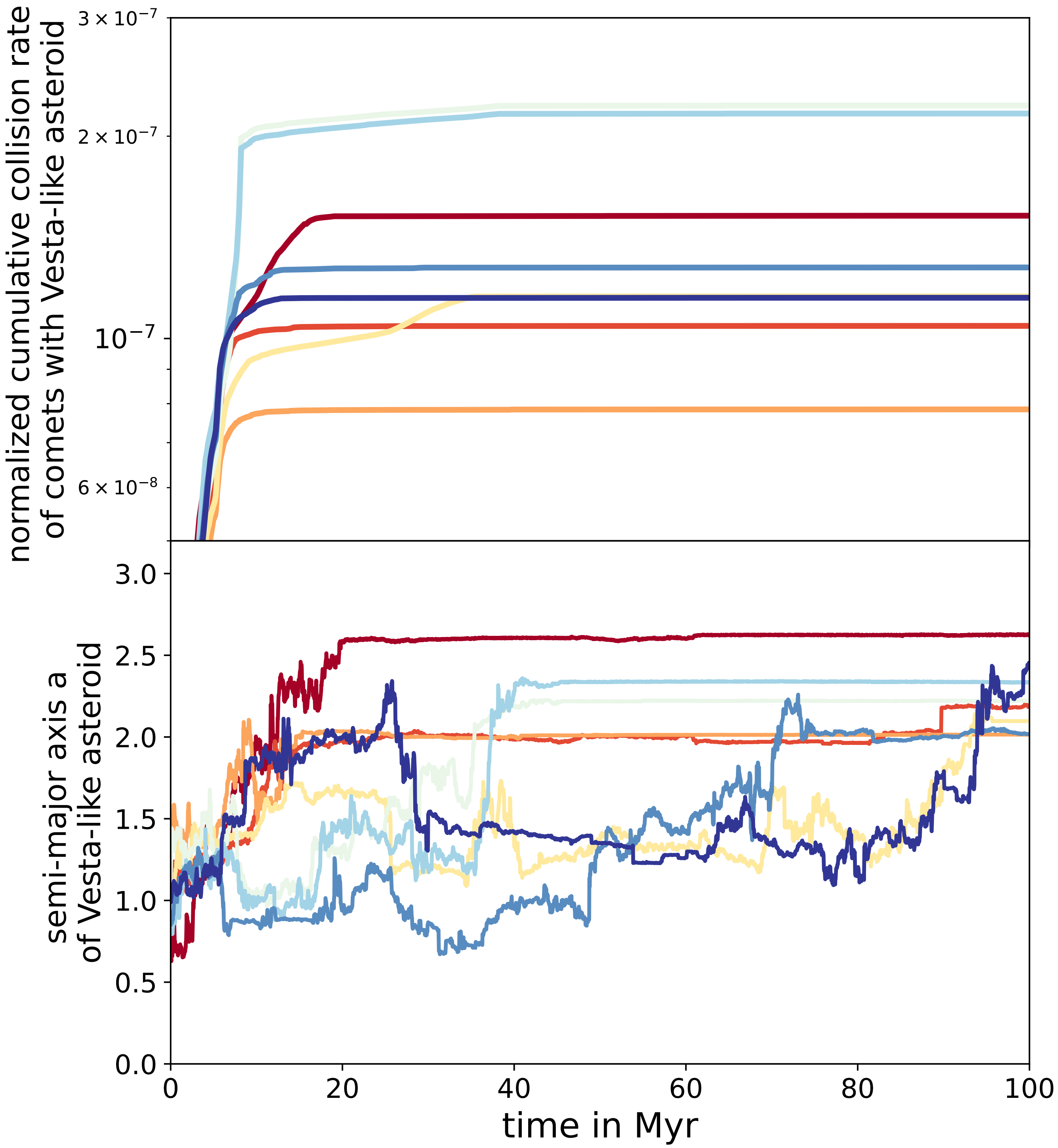

**Figure 3.** Upper panel: normalized cumulative collision rate of comets with a Vesta-like asteroid. *Normalized* means the cumulative collision rate in each simulation was divided by the total number of comets in these simulations (details of these simulations can be found in S. Joiret et al. 2024). Lower panel: semimajor axis evolution of each Vesta-like asteroid. Each Vesta-like asteroid is represented by the same color in both panels.

We tested the hypothesis that this absence of cometary signature instead results from Vesta's low escape velocity

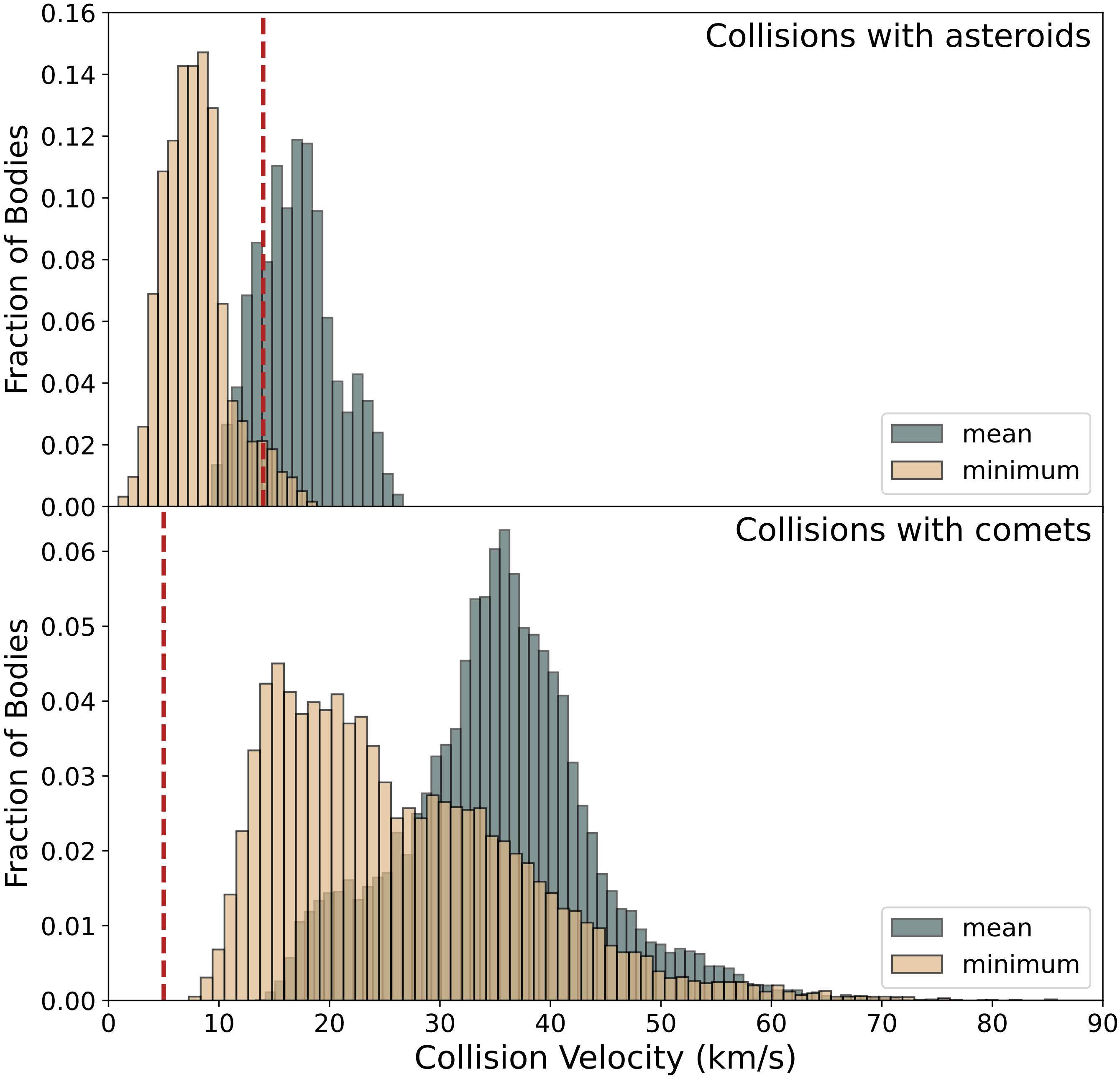

**Figure 4.** Histogram of mean and minimum collision velocities between Vesta-like asteroids and carbonaceous asteroids (upper panel) or Vesta-like asteroids and comets (bottom panel) following the giant planet instability. We only considered time steps at which there is a nonzero collision probability between the Vesta-like asteroid and carbonaceous asteroids/comets. Collision velocities between each pair of bodies are weighted by the collision probability per unit of time at each orientation. The red dashed vertical lines indicate the limit for retaining a collisional signature from asteroids and comets, respectively (see Figure 5).

($v_{esc} = 0.3632$ km s$^{-1}$), and thus also its lack of an atmosphere, which together allow only the slowest impactors to be accreted.

The lower bound for cometary collision velocities following the giant planet instability is $\simeq$10 km s$^{-1}$ (Figure 4). For each Vesta analog, at each time step of the corresponding simulation, we calculated the mean and minimum collision velocities of comets and carbonaceous asteroids on the Vesta analog. Carbonaceous asteroids typically collide with Vesta-like asteroids at substantially lower velocities than comets. Collisions with velocities as low as 10 km s$^{-1}$ would have been very rare for comets but relatively more frequent for carbonaceous asteroids.

We used a series of hydrodynamic impact simulations to evaluate how (un)likely it is for Vesta to retain material from comets, and hence a cometary signature, in such high-velocity collisions. We estimated the efficiency of material accretion depending on the velocity of cometary impactors (Figure 5). For collision velocities $\geqslant$5 km s$^{-1}$, 100% of the impactor ejecta escapes following impact. This, combined with the fact that cometary impacts with velocities <10 km s$^{-1}$ are not likely (see Figure 4), demonstrates that a cometary signature, and thus a U-Xe signature, on Vesta would be difficult to explain. Extrapolating from Vesta, which has among the highest escape velocities of all asteroids, no other asteroid is likely to have been able to retain a cometary signature.

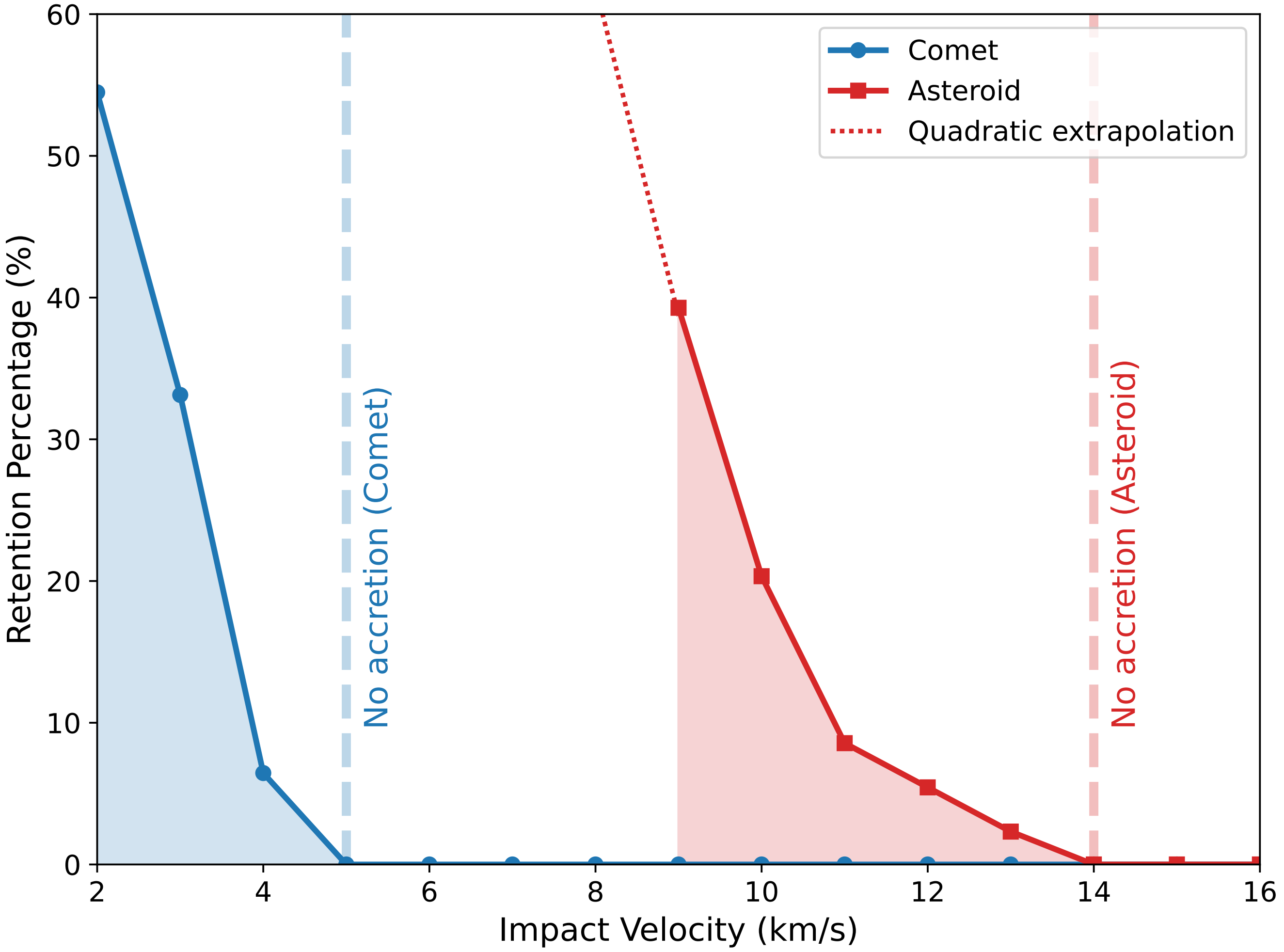

**Figure 5.** Retention percentage of impactor material on Vesta as a function of impact velocity for a 10 km comet and a 10 km asteroid. No material is retained for cometary impacts exceeding 5 km $s^{-1}$, while for asteroidal impacts, complete ejection occurs above 14 km $s^{-1}$.

It should be noted that ultracarbonaceous xenoliths, i.e., inclusions of ultracarbonaceous dust particles, have been identified within different chondrites (Y. Kebukawa et al. 2019; L. R. Nittler et al. 2019). However, these inclusions were shown to have accreted through low-velocity collisions with zodiacal cloud dust rather than via cometary impacts (A. E. Rubin & W. F. Bottke 2009; G. Briani et al. 2011, 2012). Furthermore, such inclusions are unlikely to contribute a detectable U-Xe signature to their host chondrites.

## 5. Conclusions

The diversity in cometary signatures across solar system bodies is—alongside stochastic factors (S. Joiret et al. 2023) and variable relative contributions (S. Joiret et al. 2024)—largely determined by the accretion efficiency during collisions. Accretion efficiency depends on parameters such as the escape velocity of the body, the density of its atmosphere, and the velocity of impactors.

We did not find any cometary xenon within the investigated HED meteorite samples, even though simulations indicate that collisions between Vesta and large (10–100 km) comets are highly probable. The reason for that is that material from cometary impactors with realistic velocities (i.e., $\geqslant$10 km $s^{-1}$) cannot be retained due to Vesta's weak gravitational attraction. More importantly, we predict that no asteroids could possibly retain a cometary xenon signature as a result of impacts. Thus, the detection of cometary xenon in samples returned from an asteroid by a space mission would serve as a *smoking gun*, indicating that it originated from the same source region as comets.

## Acknowledgments

We would like to thank Dr. Melissa Cashion from Purdue University for discussions about iSALE2D. Computer time for this study was partly provided by the computing facilities MCIA (Mésocentre de Calcul Intensif Aquitain) of the Université de Bordeaux and of the Université de Pau et des Pays de l'Adour, France. We also acknowledge use of the computational facilities of the Advanced Computing Research Centre, University of Bristol—http://www.bristol.ac.uk/acrc/. Numerical computations were also partly performed on the S-CAPAD/DANTE platform, IPGP, France. We acknowledge support from the CNRS MITI funding program (PhD grant to S.J., NOBLE project) and the CNRS's Programme National de Planétologie (PNP). We acknowledge funding in the framework of the Investments for the Future programme IdEx, Université de Bordeaux/RRI ORIGINS. This project has received funding from the European Research Council (ERC) under the European Union's Horizon Europe research and innovation program (grant agreement no.

101041122 to G.A.). S.J.L. acknowledges financial support from the UK National Environmental Research Council (grant No: NE/V014129/1).

## ORCID iDs

Sarah Joiret https://orcid.org/0000-0001-5985-2863
Guillaume Avice https://orcid.org/0000-0003-0962-0049
Ludovic Ferrière https://orcid.org/0000-0002-9082-6230
Zoë M. Leinhardt https://orcid.org/0000-0003-4813-7922
Simon Lock https://orcid.org/0000-0001-5365-9616
Alexandre Mechineau https://orcid.org/0009-0005-2171-1981
Sean N. Raymond https://orcid.org/0000-0001-8974-0758